\documentclass[conference]{IEEEtran}
\IEEEoverridecommandlockouts
\usepackage{cite}
\usepackage{amsmath,amssymb,amsfonts}
\usepackage{algorithmic}
\usepackage{graphicx}
\usepackage{textcomp}
\usepackage{xcolor}
\usepackage{cite}
\usepackage{amsmath,amssymb,amsfonts}
\usepackage{algorithmic}
\usepackage{graphicx}
\usepackage{caption}
\usepackage{textcomp}
\usepackage{xcolor}
\usepackage{svg}
\usepackage{afterpage}
\usepackage{booktabs}
\usepackage{float}
\usepackage{subfigure}
\usepackage{multirow}
\usepackage{todonotes}
\usepackage{soul}
\usepackage{lipsum}
\usepackage{graphicx}
\renewcommand{\baselinestretch}{1}
\usepackage[citecolor=blue,colorlinks=true,linkcolor=blue]{hyperref}%
\def\BibTeX{{\rm B\kern-.05em{\sc i\kern-.025em b}\kern-.08em
    T\kern-.1667em\lower.7ex\hbox{E}\kern-.125emX}}
\begin{document}

\title{Quantised Neural Network Accelerators for Low-Power IDS in Automotive Networks}

\author{\IEEEauthorblockN{Shashwat Khandelwal, Anneliese Walsh \& Shanker Shreejith}
\IEEEauthorblockA{ Reconfigurable Computing Systems Lab, Electronic \& Electrical Engineering\\
Trinity College Dublin, Ireland\\
Email: \{khandels, walsh61, shankers\}@tcd.ie}}



\maketitle

\begin{abstract}
In this paper, we explore low-power custom quantised Multi-Layer Perceptrons (MLPs) as an Intrusion Detection System (IDS) for automotive controller area network (CAN). 
We utilise the FINN framework from AMD/Xilinx to quantise, train and generate hardware IP of our MLP to detect denial of service (DoS) and fuzzying attacks on CAN network, using ZCU104 (XCZU7EV) FPGA as our target ECU architecture with integrated IDS capabilities. 
Our approach achieves significant improvements in latency (0.12\,ms per-message processing latency) and inference energy consumption (0.25\,mJ per inference) while achieving similar classification performance as state-of-the-art approaches in the literature. 
\end{abstract}

\section{Introduction \& Proposed Architecture}\label{sec:introduction}
Controller Area Network (CAN) continues to be the most widely used network architecture for in-vehicle networks owing to its cost-effective nature and ease of use in control applications. 
However, CAN (and other automotive networks like FlexRay, LIN) does not offer any inherent security and authentication mechanisms, which has been exploited in numerous attacks on in-vehicle systems~\cite{miller2015remote}. 
Intrusion detection approaches are aimed at detecting such threats on the network and allowing critical systems to enter into a `safe working' mode when such threats are detected. 
Machine Learning (ML) approaches have shown significant improvement in detection accuracy of such threats and ability to adapt to newer attack vectors~\cite{ma2022gru, desta2020mlids, song2020vehicle}; however, integrating them within a vehicular network is still challenging due to dedicated ECU integration (higher power, space, wiring) with accelerators like GPUs (to meet inference latency). 

\begin{figure*}
\centering
  \includegraphics[width=\textwidth, width=16cm,height = 5.65cm]{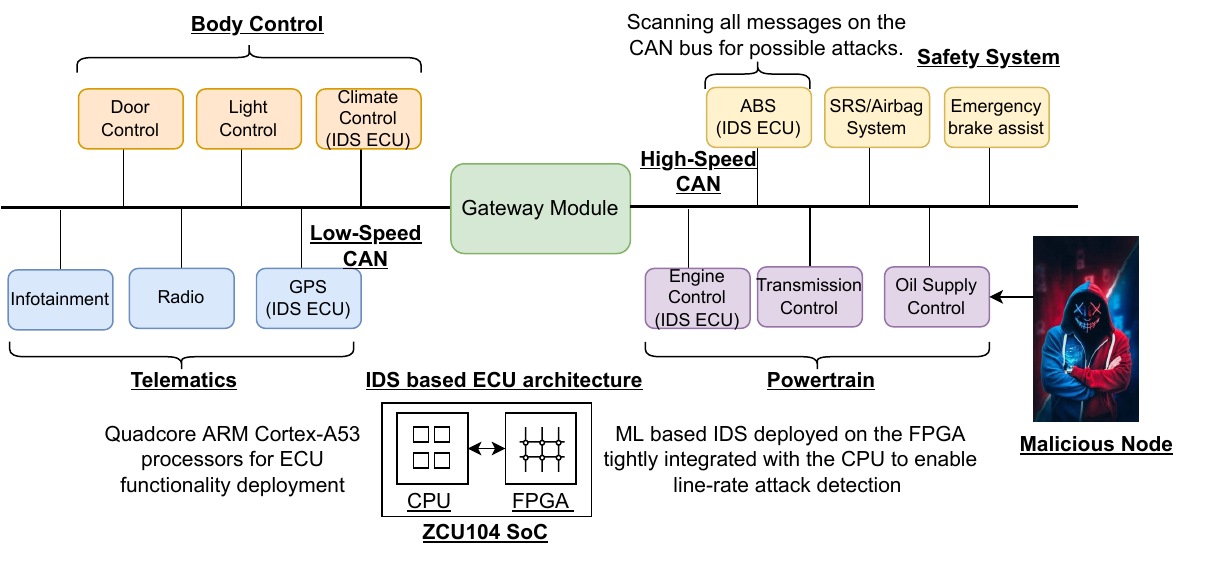}
  \caption{System diagram of the CAN network (featuring components connected to high/low speed CAN) incorporating ECUs capable of intrusion detection.}
  \label{fig:datapath}
\end{figure*}

In this work, we investigate a custom quantised multi-layer perceptron (MLP) model as an IDS and its integration as a coupled accelerator on an ECU.
We have explored quantised Convolutional Neural Network (CNN) models as IDSs in our previous works but found lower classification performance when the whole frame is used as the input feature~\cite{khandelwal2022deep,khandelwal2022light}.
The proposed architecture, shown in fig.~\ref{fig:datapath}, uses a AMD/Xilinx Zynq Ultrascale+ device acting as a standard ECU, with our IDS model integrated as a custom IP on the programmable logic. 
In our approach, CAN packets received in the interface are handled as usual by the ECU to perform its task; additionally, the packet is copied into a FIFO style buffer capturing a time-series of messages, which is examined by our IDS IP for threat signatures. 
This integration uses the ECU to control and manage the IDS accelerator through the software APIs, making them similar to accelerators enabled through AUTOSAR compliant abstraction layers. 
The MLP is defined using PyTorch and undergoes quantisation aware training using AMD/Xilinx's FINN framework~\cite{umuroglu2017finn}.
We utilise the openly available CAR Hacking dataset~\cite{song2020vehicle} for training and validation, which provides a labelled set of real CAN messages and multiple attack vectors captured from the OBD port of an actual vehicle.
We train two models, to detect DoS attacks and Fuzzying attacks, using the Brevitas quantisation aware training library from AMD/Xilinx~\cite{brevitas}.
Design space exploration is performed to arrive at the quantisation level to reduce the resource consumption and computational complexity without compromising on the detection accuracy. 
From our experiments, we observed that 4-bit uniform quantisation achieved best performance in both DoS and Fuzzying attacks, and hence was chosen for deployment.
Streaming layer optimisations and partitioning were chosen during FINN compilation flow to optimise the hardware IP generated from the trained MLP model. 
Subsequently, the generated IP is integrated into a standard Zynq design flow as a slave memory-mapped peripheral device.
The FINN flow also generates low-level drivers for interfacing our IP from the ARM cores on the Zynq device over the AXI interface. 
We use a Linux operating system (from the PYNQ image) with low-level Xilinx run-time tools integrated to use these APIs for testing our approach. 
For our tests, we integrate a single IP core (DoS or Fuzzying) to test their performance in isolation.

\section{Results \& Conclusion}\label{sec:experiments}
To evaluate the proposed approach as an integrated IDS accelerator, we evaluate the classification performance of the model in identifying DoS and Fuzzy attacks from the Car Hacking dataset and compare them against state of the art techniques in the literature.
Table~\ref{table:comp1} compares the performance of both our models (4-bit QMLP) against other results from the literature, in terms of inference precision, recall, F1 score and false negative rate. 
Subsequently, we quantify the per-message processing latency of the model, starting from the arrival of the CAN message at the interface to determine the detection delay incurred by the approach.
Table~\ref{table:latcomp} compares our result against other approaches in literature, which utilise different platforms (GPUs, Jetson edge accelerators, Rasperry Pi) and approaches (block of CAN messages v/s individual messages). 
It should be noted that the latency metric in the case of block-based IDS does not consider the delay in acquiring the number of CAN messages required, and hence could result in potentially larger delays in attack detection.
Our approach achieves 0.12\,ms per CAN frame, which is a 4.8$\times$ improvement over the dedicated line-rate MTH-IDS approach~\cite{yang2021mth}.
In terms of throughput, our QMLP coupled ECU can process over 8300 messages per second at highest payload capacity, achieving near-line-rate detection on high-speed critical CAN networks. 
We observe that our model consumed 2.09\,W when measured directly from the device's power rails (using the PYNQ-PMBus package) while performing inference and other tasks on the ECU (with Linux OS), thus consuming 0.25\,mJ of energy per inference, significantly better than other approaches that use GPUs (9.12\,J per inference for our 8-bit quantised MLP model on an A6000 for reference) and/or dedicated CPUs for IDS. 
The single model deployed consumes less than 4\% of resources on the device, allowing multiple models to be executed simultaneously for a comprehensive IDS integration at slightly higher energy consumption than what we observed, making it an ideal deployment framework for line-rate IDS in automotive networks. 

\begin{table}[t!]
\centering
\caption{Accuracy metric comparison (\%) of our quantised FPGA accelerators against the IDSs in reported literature.}
\scalebox{1}{
\begin{tabular}{@{}llllll@{}}
\toprule
\textbf{Attack}  & \textbf{Model} & \textbf{Precision} & \textbf{Recall} & \textbf{F1}  & \textbf{FNR} \\
\midrule
\multirow{5}{*}{DoS} 
& DCNN~\cite{song2020vehicle}                  & 100                & 99.89          & 99.95  & 0.13     \\
& MLIDS~\cite{desta2020mlids}                  & 99.9                & 100          & 99.9  & -     \\
& NovelADS~\cite{agrawal2022novelads}                  & 99.97               & 99.91          & 99.94  & -     \\
& TCAN-IDS~\cite{cheng2022tcan}                  & 100             & 99.97          & 99.98  & -     \\
& GRU~\cite{ma2022gru}                  & 99.93             & 99.91               & 99.92  & -     \\
& \textbf{4-bit-QMLP}                  & 99.99             & 99.99               & 99.99  & 0.01     \\
\midrule
\multirow{5}{*}{Fuzzy} 
& DCNN~\cite{song2020vehicle}                 & 99.95             & 99.65          & 99.80  & 0.5     \\
& MLIDS~\cite{desta2020mlids}                  & 99.9             & 99.9          & 99.9  & -     \\
& NovelADS~\cite{agrawal2022novelads}                  & 99.99               & 100         & 100  & -     \\
& TCAN-IDS~\cite{cheng2022tcan}                  & 99.96             & 99.89          & 99.22  & -     \\
& GRU~\cite{ma2022gru}                  & 99.32             & 99.13               & 99.22  & -     \\
& \textbf{4-bit-QMLP}            & 99.68             & 99.93          & 99.80   &  0.07    \\
\bottomrule
\end{tabular}}
\label{table:comp1}
\end{table}

\begin{table}[t!]
\centering
\caption{Per-message latency comparison against other state-of-the-art IDSs reported in literature.}
\scalebox{1}{
\begin{tabular}{@{}lrll@{}}
\toprule
\textbf{Models}     & \textbf{Latency} & \textbf{Frames} & \textbf{Platform} \\ \cmidrule{1-4}
GRU~\cite{ma2022gru} & 890\,ms & 5000 CAN frames & Jetson Xavier NX  \\
MLIDS~\cite{desta2020mlids} & 275\,ms & per CAN frame & GTX Titan X \\
NovelADS~\cite{agrawal2022novelads} & 128.7\,ms & 100 CAN frames & Jetson Nano \\
DCNN~\cite{song2020vehicle} &  5\,ms & 29 CAN frames & Tesla K80 \\
TCAN-IDS~\cite{cheng2022tcan} & 3.4\,ms & 64 CAN frames & Jetson AGX \\
MTH-IDS~\cite{yang2021mth} &  0.574\,ms & per CAN frame & Raspberry Pi 3     \\
\textbf{4-bit-QMLP} (ours) & 0.12 ms & per CAN frame & Zynq Ultrascale+    \\ 
\bottomrule
\end{tabular}}
\label{table:latcomp}\vspace{-5mm}
\end{table}


\bibliography{references}
\bibliographystyle{ieeetr}

\end{document}